\documentclass[twocolumn]{article} 
\usepackage{graphicx}
\usepackage{amsmath, amssymb}
\usepackage{caption}
\usepackage{cite}
\usepackage[a4paper]{geometry}
\usepackage{abstract}
\usepackage{dblfloatfix} 
\usepackage[dvipsnames]{xcolor}
\usepackage[inkscapelatex=false]{svg}

\title{Nanomechanical Error Correction}

\author{%
Xiaoya Jin$^{*}$, 
Christopher G. Baker, 
Erick Romero, 
Nishta Arora,
Nicolas P. Mauranyapin, \\
Timothy M. F. Hirsch, 
Glen I. Harris, 
Warwick P. Bowen$^{\dagger}$\\
 \\
School of Mathematics and Physics, The University of Queensland, QLD 4072, Australia
}

\date{} 

\begin{document}

\twocolumn[

\maketitle

\begin{center}
\textbf{Abstract}
\end{center}
\noindent

Error correction is essential for modern computing systems, enabling information to be processed accurately even in the presence of noise. Here, we demonstrate a new approach which exploits an \textit{error correcting phase} that emerges in a system of three coupled nonlinear resonators. Within this phase, perturbed memory states are autonomously restored via the collective dynamics of the nonlinear network. We implement our scheme using a network of nanomechanical resonators. Nanomechanical systems are an attractive platform for low energy computing, but purely mechanical error correction has not been previously demonstrated. We experimentally show that the error correcting phase provides a 35 times reduction in the rate of errors, and allows robust error correction over a wide range of system parameters. These results highlight how emergent nonlinear dynamics can be harnessed for practical applications, paving the way towards error-resilient nanomechanical computing.

\vspace{2cm} 
]

\begingroup
\renewcommand{\thefootnote}{\fnsymbol{footnote}} 
\footnotetext[1]{Email: \texttt{tina.jin@uq.edu.au}}
\footnotetext[2]{Email: \texttt{w.bowen@uq.edu.au}}
\endgroup

\setcounter{footnote}{0}

\section*{Introduction}
\textbf{}

Emergent collective dynamics — complex, often unexpected, macroscopic behaviours that arise from local interactions — are a defining feature of driven, dissipative, coupled nonlinear systems. They can be observed in both engineered systems and nature, with examples including the synchronisation of coupled oscillators \cite{Bagheri_13_PRL, Doster_19_NatureComm}, phase transitions in spin systems \cite{Minganti_18_PRA}, oscillatory chemical reactions \cite{Wood_85_J.ChemPhys}, and pattern formations within vegetation growth \cite{Lejeune_04_IJQC}. These systems highlight how the complex interplay of energy gain, loss, and nonlinearity can give rise to rich dynamical behavior when far from equilibrium, often with properties not apparent from the behavior of individual components alone. An improved understanding of such dynamics has inspired several technological innovations, including the use of ferromagnetic interactions to stabilise bit storage within magnetic storage devices \cite{Daughton_97_JAP}, and coherent Ising machines that exploit nonlinear optical interactions to realise collective optimisation \cite{McMahon_16_Science}.

Recent advances in nanofabrication have also enabled the study of dissipative, nonlinear dynamics in nanomechanical systems \cite{Cleland_96_APL, Roukes_01_world, Schmid}. For example, it has been demonstrated that a system of eight ring-coupled nonlinear nanomechanical resonators exhibits exotic states of synchronisation and symmetry breaking \cite{Roukes_science_2019}. These systems exhibit readily accessible nonlinear responses and controllable oscillation properties, rendering them ideal for studies into nonlinear phenomena, as well as for practical applications \cite{Rieger_12_APL, Eichler_13_nature_comm, Leuch_16_PRL, Seitner_17_PRL, Moser_13_nature_nanotech, ChasteJ.2012Anms, GavartinEmanuel2013Soal, HanayM.S.2012Snms}. In particular, nanomechanical systems are a versatile and attractive platform for performing computation \cite{Yasuda_nature_21}.

In a mechanical computer, binary information can be encoded in the amplitude and or phase of the vibrations of nanomechanical nonlinear resonators \cite{Roukes_IEEE_04}, playing a role analogous to electrical charges in semiconductor devices. Compared to electronics, nanomechanical components promise longer lifespans in harsh environments and ultra-low power consumption, in principle approaching the Landauer limit \cite{Bennett_82_IJTP, Lee_NatComm_23, Romero_PRA_24}. This is particularly desirable in the current computing landscape, with artificial intelligence models consuming increasing amounts of energy for both training and interfacing \cite{Kim_25_NatureNanotech}.

While significant advances have been made towards realisation of a nanomechanical computer \cite{Wenzler_Nano_Lett_14, Yamaguchi_nat_comm_11, Ilyas_2019, Hatanaka_Nature_17, Tadokoro21, Badzey_APL_04, Bagheri_nat_nano_11, Li_Sci_Adv_24, Song, Mahboob_14_SciRep, Dion_18_JApplPhys, Dubcek_24_IPSC}, purely mechanical error correction (which operates entirely in the mechanical domain without converting information into electrical signals) has yet to be demonstrated. It has recently been proposed that mechanical error correction can be achieved by harnessing an error correcting phase that emerges from the collective dynamics of coupled nonlinear resonators \cite{Jin_SciRep_24}. Here, we experimentally demonstrate the spontaneous emergence of this error correcting phase in a system of three tunable nonlinear nanomechanical resonators. Within the phase, perturbed memory states are autonomously restored to their correct values and performance is robust against experimental imperfections. We observe that the error correcting phase reduces errors by a factor of 35. Our work shows how emergent nonlinear dynamics can be harnessed for practical applications, opening a path towards scalable, error-resilient nanomechanical computing.

\begin{figure*}[!b]
    \begin{center}
    \includegraphics[width= 1 \linewidth]{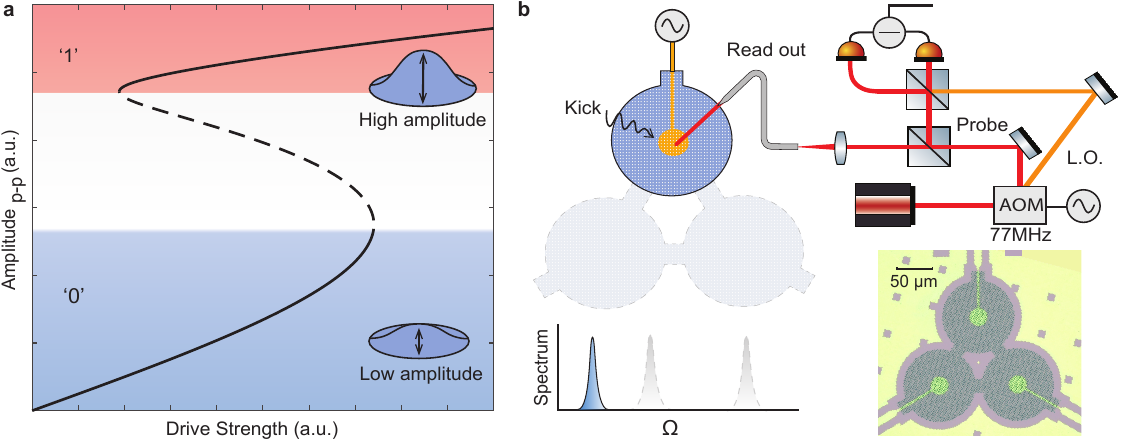}
    \end{center}
    \caption{\textbf{Concept and implementation of a single nanomechanical memory unit.} \textbf{a} Drive response of a Duffing resonator \cite{Schmid, Romero_PRA_24}. At a sufficiently large drive strength, multiple solutions of the Duffing equation emerge, with the stable and unstable solutions represented by solid and dashed lines, respectively. Blue and red regions represent `0' and `1' logic bits, respectively. \textbf{b} Experimental setup for a system of three resonators. Optical heterodyne interferometry is used to measure the amplitude of the resonator. The laser light first passes through an acousto-optic modulator (AOM), which generates a frequency upshifted ($77$ MHz) local oscillator (LO) beam (orange), and a non-diffracted probe beam (red). The probe beam is directed to the vacuum chamber via a fiber optic cable. A lensed fiber focuses the light onto the membrane and collects the reflected light.  An optical microscope image of the fabricated device is shown in the bottom right. An illustrated mechanical power spectrum of the system is shown in the bottom left, showing three distinct resonances, corresponding to the fundamental mode of each of the uncoupled resonators.}
\end{figure*}

\section*{Results}

The nonlinear resonators used in this work are Duffing resonators, which are bistable under certain drive strengths and frequencies, exhibiting two steady state oscillation amplitudes \cite{Schmid,Lifshitz_08}, as illustrated in Fig. 1a. Above a critical threshold, $F_\text{crit}$, the resonator will be driven into one of two stable limit cycles. Binary memory can be encoded in the oscillation amplitude \cite{Romero_PRA_24, Tadokoro21}, with high amplitude representing a binary `1' and low amplitude representing a binary `0'. If exposed to an impulse, henceforth referred to as a `kick', the resonator can dynamically transition from one stable branch to the other, resulting in a bit flip known as a single event upset. The error correcting system proposed in Ref. \cite{Jin_SciRep_24} overcomes this by exploiting the nonlinear response of Duffing resonators when kicked. This results in a complex interaction between the coupled resonators and the external drive, enabling the kicked resonator to more rapidly dissipate its excess energy and automatically return to its original state.

\subsection*{Architecture}

To experimentally test this prediction, we devised a physical architecture consisting of three circular membrane resonators as shown in the optical image in Fig. 1b (for fabrication details, see Methods). The resonators are formed by underetching highly stressed silicon nitride membranes, and exhibit drum-like resonances at a nominal resonant frequency ($\Omega_0/2\pi$) of 4 MHz with quality factors ($Q = \Omega_0/\Gamma$) of approximately 26,000, where $\Gamma/2\pi$ is the linewidth. Gold electrodes deposited at the centre of each resonator allow the devices to be capacitively actuated \cite{Romero_PRA_24}. The resonators are acoustically connected in a ring via acoustic tunnel junctions \cite{Mauranyapin_2021_PRA}, which permit the evanescent transfer of mechanical energy. The mechanical motion of each device is detected optically \cite{Romero_PRA_24, Mauranyapin_2021_PRA, Hirsch_24_APL}, as shown in Fig. 1b. Laser light is focused onto the membrane via a lensed optical fiber positioned above it. Incident light is reflected by the membrane, with its phase dependent on the position of the membrane. The amplitude of the motion of the resonator can then be measured using a heterodyne interferometer (see Methods).

The resonant frequencies of the resonators were found to vary by up to 0.5$\%$ of $\Omega_0$ ($\sim 20$ kHz) due to minor fabrication differences. Because the linewidth ($\Gamma/2\pi \approx 175$ Hz) is significantly less than this spread, the resonators are natively uncoupled, as illustrated in the schematic spectrum in Fig. 1b. DC voltages can be applied to the electrodes to tune the frequencies via the electrostatic softening effect \cite{Schmid, Kozinsky_PhysicsApplied_2006}. This allows us to induce strong coupling between the resonators and create a system of three all-to-all coupled resonators. Toggling the DC voltages enables us to examine both isolated resonators and the coupled system using the one physical setup.

\subsection*{Single Resonator}

To understand the dynamics of a single upset event, we first characterise the effect of kicks on a single unprotected nanomechanical memory bit, before comparing to experiments on the coupled system. 

To study an isolated resonator,  we drive one resonator \textemdash{} uncoupled to the others \textemdash{} to reach the bistable regime using a DC signal and an AC signal near its resonant frequency (see Methods). Simultaneously, we optically measure the resonator's response, as illustrated in Fig. 1b. We apply kicks to the resonator to induce bit flip errors by superimposing a short sinusoidal pulse (at the drive frequency) on top of the existing drive (see Supplementary Information). The kick signal applies a large force on the resonator ($F_\text{kick}/ F_\text{crit} \approx 125$) over a short period of time ($T_\text{kick}\approx 12\ \mu $s), such that it is `instantaneous' compared to the characteristic response time of the resonators ($\tau = 2\pi/\Gamma \approx 6$ ms). Indeed, the finite kick duration represents less than $0.2 \%$ of the characteristic response time of the resonator ($T_\text{kick}/\tau < 0.002$), such that it accurately models a single event upset \cite{Wang_17_SEU}.

\begin{figure*}[!ht]
\begin{center}
\includegraphics[width= 0.975 \linewidth]{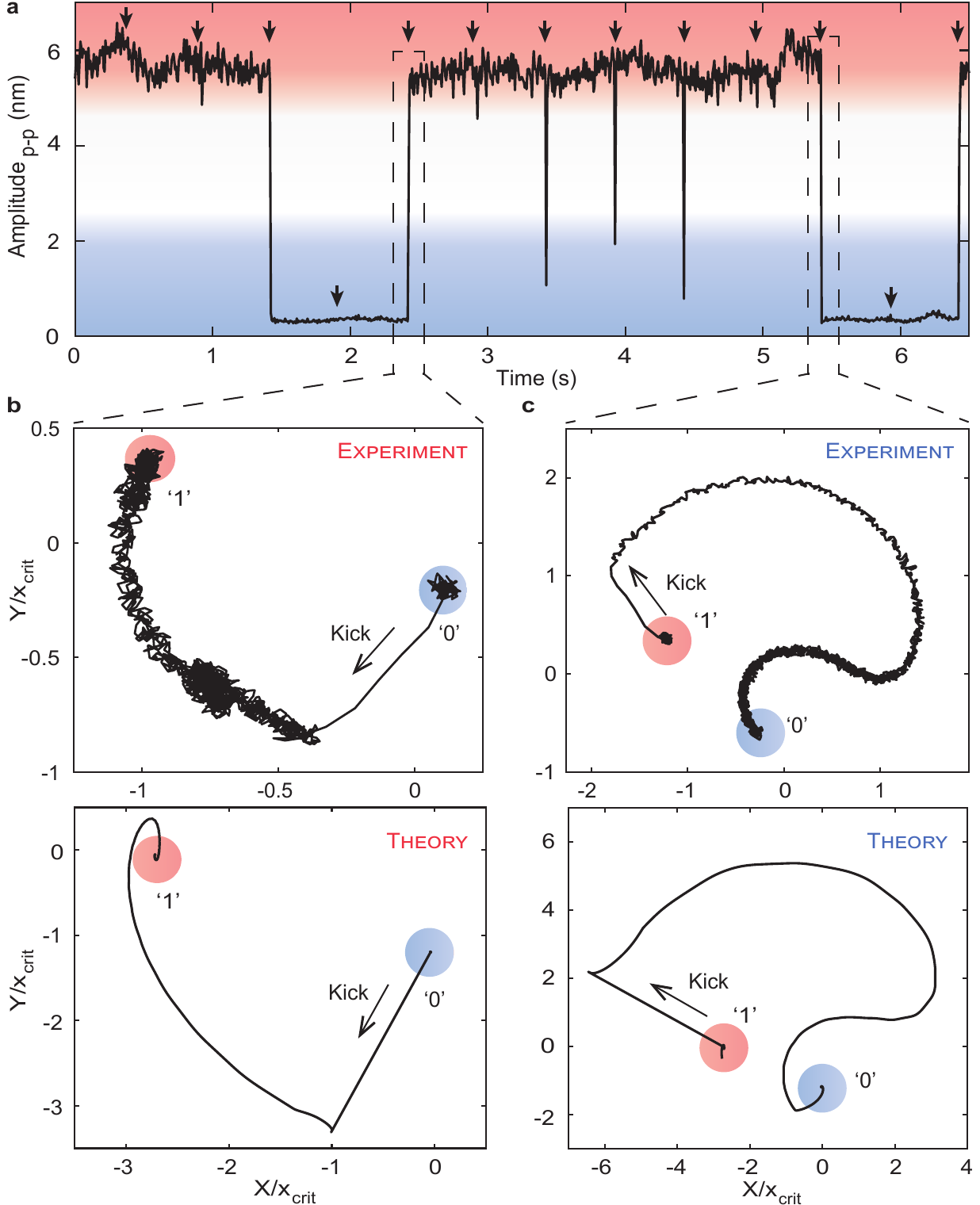}
\end{center}
\caption{\textbf{Observed bit flip errors from a single, isolated resonator.} \textbf{a} An amplitude-time measurement of the isolated resonator. Black arrows indicate the timing of kicks. The background is shaded (blue/red) to indicate the regions where the oscillation amplitude is considered to be in the `0'/`1' state. \textbf{b/c} Experimental (top) and simulated (bottom) phase trajectories, respectively showing example `0' to `1' and `1' to `0' bit flip dynamics. Red and blue circles are respectively used to illustrate the resonator's `1' and `0' states. The axes are normalised by the critical amplitude, x$_\text{crit} \approx 4.0$ nm, which is the peak-to-peak amplitude of the resonator when it first exhibits nonlinear behaviour. The experimental traces were measured over 3 ms, and the theoretical traces simulated over 10 $\mu$s (simulation details provided in Methods). }
\end{figure*}

Fig. 2a shows an example measured time trace of the peak-to-peak oscillation amplitude of the isolated resonator, to which kicks are applied every 0.5 seconds. We observe that four of the thirteen kicks give rise to sudden and sustained transitions between the `1' (high amplitude) and the `0' (low amplitude) states. These are bit flip errors, which would result in information loss or corruption in a large-scale nanomechanical computer. The effect of kicks on the resonator varies because of the randomised relative phase between kicks and resonator motion, leading some kicks to oppose the momentum of the resonator at the time of incidence and others to add to it \cite{Jin_SciRep_24}. An error correcting system must be able to prevent bit flip errors regardless of the kick timing.

To investigate the dynamics of bit flip errors, we track the trajectory of the resonator in phase space in a frame rotating with the input sinusoidal drive during an upset event. This is achieved by mixing down the recorded photocurrent with two sinusoids out of phase by $\pi/2$ (see Methods). With sufficient signal to noise ratio, we can extract the amplitude and phase of the resonator in a time shorter than the mechanical period ($T = 2 \pi /\Omega_0$), enabling us to track not just the timing of the errors (Fig. 2a), but also the precise evolution of the resonator in phase space as the kick forces a transition from one limit cycle to another. The resultant phase trajectories are shown in Fig. 2b. In the rotating frame, the two stable limit cycles manifest as single points in phase space.

Fig. 2b shows the dynamics of a `0' to `1' bit flip error, recorded over $2.5$ ms. Here, the resonator is initially in the `0' state (blue circle). The impulse from the kick causes a sharp increase in peak-to-peak amplitude, dislodging the resonator from the stable `0' limit cycle. The resonator does not recover, and transitions into the `1' state (red circle). Similarly, Fig. 2c shows the phase space trajectory of a `1' to `0' bit flip error, where the same progression occurs but with the initial and final states interchanged.

To validate the observed error trajectories, we simulate the dynamics by numerically solving the driven, damped Duffing equation (see Methods). The large difference between decay and oscillation timescales at our experimental quality factor made accurate simulations difficult. Instead, the simulations were performed using a lower quality factor ($Q = 10$). This is justified since the evolution of the system is governed primarily by the position and momentum envelopes, rather than the fast oscillation. We adjust the timing of the kick to mimic the phase-angle observed in experiment. The simulation results, in an equivalent rotating frame, are shown in the lower panels of Fig. 2b and 2c. We observe strong qualitative agreement with experiment, with scaled coordinates and quadratures.

\subsection*{Coupled Resonators}

To now test the error correction ability of three coupled resonators, we apply DC voltages to bring the disparate resonant frequencies together. Figure 3a shows the resonant frequency of the resonator with the highest natural frequency (4.143 MHz) as DC voltages are applied. As the DC voltage is progressively increased, the resonator's resonant frequency decreases. As it crosses the natural frequencies of the other two resonators (4.141 MHz and 4.132 MHz), we observe clear anti-crossings (see also Fig. 3a inset), evidencing strong coupling. The simulated power spectrum shown in the background shows excellent agreement with experiment (see Methods).

\begin{figure*}[!h]  
    \centering
    \includegraphics[width=1\linewidth]{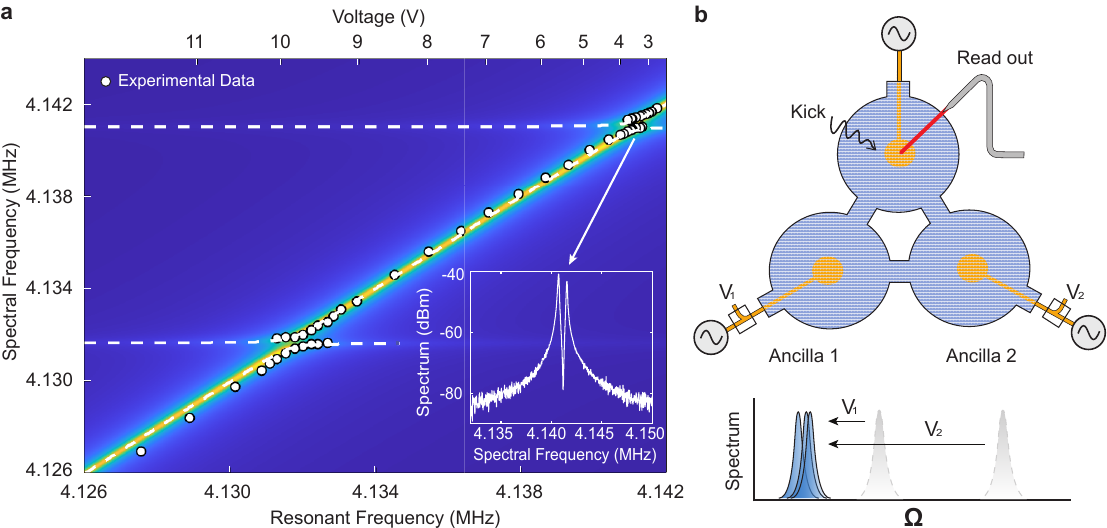}
    \caption{\textbf{Mechanism and setup for resonator coupling.} \textbf{a} Power spectrum of the highest frequency resonator with varying applied DC voltage. Here, the \textit{read out} resonator is driven with AC and DC voltages, while the two other resonators are not driven. The theoretical power spectrum is plotted in the background (see Methods), with the experimental data (white dots) overlaid. White dashed lines are plotted to indicate anti-crossings in the spectrum at the frequencies of the two other resonators. \textbf{b} Experimental setup for all-to-all coupling. Independent DC voltages are applied to \textit{ancilla} resonators 1 and 2 to create a system of three, all-to-all coupled resonators. The spectrum of the system is shown in the lower illustration, showing three overlapping resonances.}
    \label{fig:coupled_schematic}
\end{figure*}

By applying independent DC voltages to the two higher frequency resonators, we tune them to be degenerate with the lowest frequency resonator, which is subject to a comparably smaller DC signal. This gives rise to a system of three all-to-all coupled resonators, as illustrated in Fig. 3b. In this system, the three fundamental drum modes hybridise to form three new normal modes (see Supplementary Information). The lowest frequency normal mode exhibits symmetric, equal-amplitude, in-phase vibrations across the three resonators. It thus forms a natural mode for storing three identical memory bits (`000' or `111'). Importantly, the error correction should not be thought of simply in terms of a single vibrational normal mode. When a kick occurs, the Duffing nonlinearity of the kicked resonator causes its frequency to shift. This alters the normal modes \cite{Jin_SciRep_24} of the system, and is essential for the error correcting process. To drive the mode, we actuate all three resonators near resonance with a synchronised amplitude and phase. Henceforth, we measure the dynamics of one resonator, referred to as the \textit{read out} resonator. We refer to the remaining resonators as the \textit{ancilla} resonators.

For comparison purposes, we apply the same experimental protocol used for the single resonator to induce kicks to the coupled system. That is, the kicks have the same magnitude as previous tests and they are only applied to the read out resonator. Two of the resultant experimental time traces are shown in Fig. 4a and 4b, with the coupled system initialised in the `0' and `1' state, respectively. In both traces, kicks can cause the amplitude of the read out resonator to be momentarily upset, but the resonator consistently returns to its original logical state.

The top panels of Fig. 4c and 4d show the phase space trajectories for a kick from each of the traces, displayed over $5$ ms. We observe that immediately after the kick, the read out resonator's trajectory sharply deviates from the `0'/`1' steady state limit cycle and then `loops around' before returning to its original state. Qualitatively similar behaviour is observed in the simulated response (see Methods for simulation details), shown in the lower panels of Fig. 4c and 4d. 

Notably, the error correction mechanism is not simply due to the increased effective mass and characteristic energy (and therefore resistance to kicks) of the three resonator normal mode. The error correction achieved by the system only arises due to a complex interaction between the coupled resonators and with the external drive \cite{Jin_SciRep_24}.  When the amplitude of a Duffing resonator is altered, its nonlinear nature causes its frequency to shift. The kick to the read out resonator thus detunes it from the ancilla resonators, such that it is dynamically decoupled and unable to exchange energy efficiently. This frequency shift, combined with the sudden changes in amplitude and velocity, also causes the read out resonator to rephase its oscillation with respect to the external drive; with the resulting transfer of energy favoring a return to the initial state.

To verify this dynamical behaviour, we simulated the power flows from the external drive to the read out resonator and the intrinsic dissipation with the system initialised in the `0' and the `1' state, as shown in Fig. \ref{error correction mechanism}a and 5b, respectively. Prior to the kick, the external drive provides enough energy to overcome dissipation and maintain the oscillation. The kick in Fig. \ref{error correction mechanism}a increases the energy of the resonator, but also changes the relative phase between the resonator and the drive by more than $\pi/2$, such that the direction of power flow is reversed. This mechanism results in energy being removed from the system faster than would occur via intrinsic losses alone (red curve), assisting the system in automatically returning to its original state. Where the kick reduces the energy of the read out resonator, as in Fig. \ref{error correction mechanism}b, the read out resonator instead gains energy from the external drive via rephasing. These periods of rephasing manifest as `looping' cycles in the phase space trajectories, as observed in Fig. 4c and 4d.

\begin{figure*}[!ht]
\begin{center}
\includegraphics[width= 1 \linewidth]{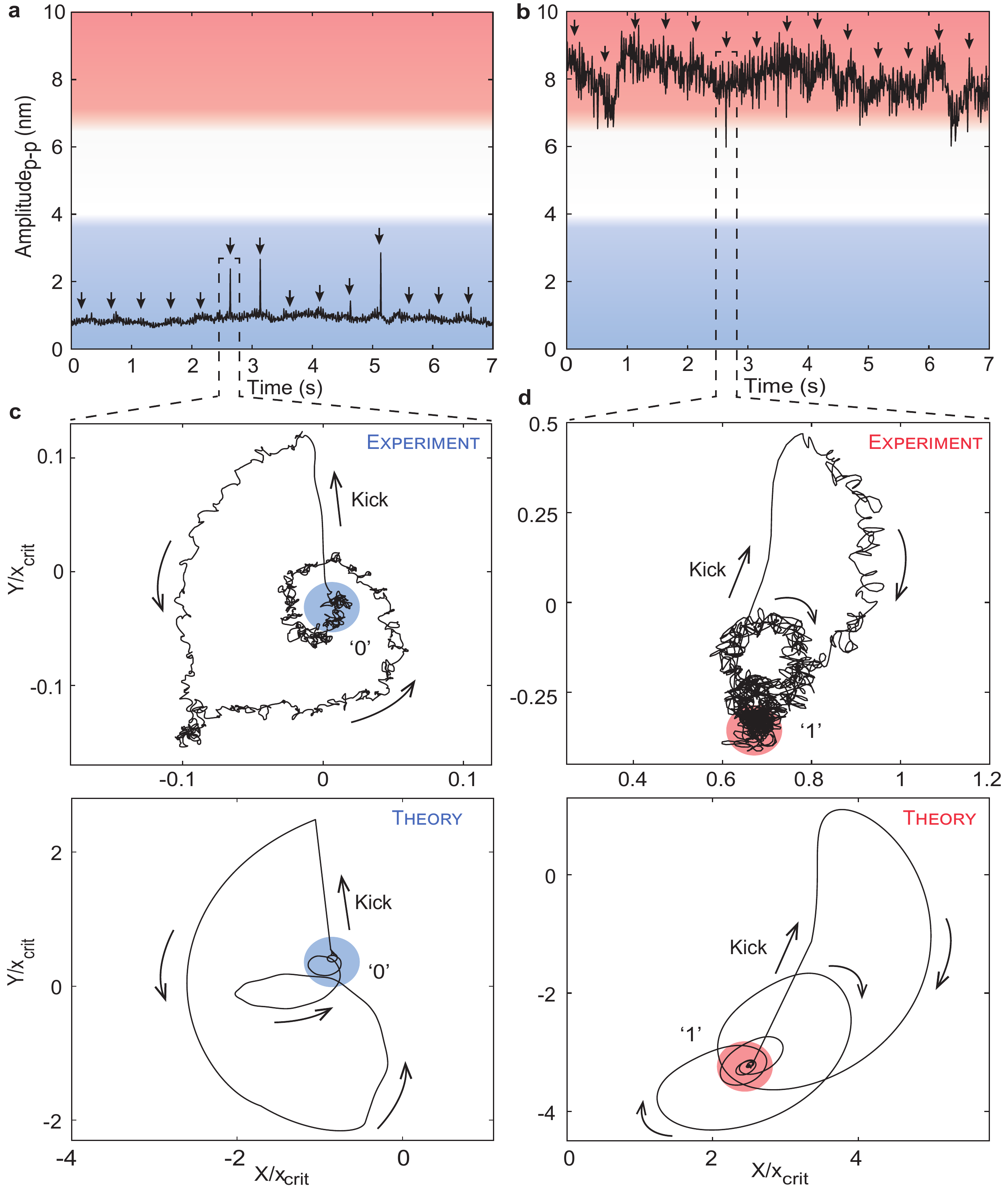}
\end{center}
\caption{\textbf{Autonomous nanomechanical error correcting memory.} \textbf{a/b} Amplitude-time measurement of the coupled resonators when initialised in their `0' (left) /`1' (right) state. Black arrows are used to indicate when the kicks are applied. \textbf{c/d} Experimental (top) and simulated (bottom) error correcting dynamics of three coupled resonators. Phase space trajectory axes are normalised by the critical amplitude (x$_\text{crit} \approx 4.0$ nm). Arrows are used to indicate the direction of the trajectory. The experimental trajectories are measured over 3 ms, and the simulated trajectories over $55\ \mu$s.  }

\label{coupled kick dynamics}
\end{figure*}

To test how significant the dynamic decoupling effect is in our system, we track the frequency shift of the read out resonator in response to a kick, as shown in Fig. 5c. We observe that the applied kick results in an instantaneous frequency shift of up to 0.5 linewidths, temporarily reducing the efficiency of energy exchange between resonators. For even larger kicks, the read out resonator's frequency would be shifted by more than a linewidth, causing it to become fully dynamically decoupled from the ancilla resonators. In this regime, energy exchange between the resonators is greatly suppressed, protecting the ancilla resonators from the excess energy introduced by the kick, and allowing the read out resonator to dissipate this energy. Eventually, the amplitude of the read out resonator would reduce sufficiently for coupling to be re-established and the system returns to its initial state. In this way, the error correction scheme is predicted to be robust even to very large kicks \cite{Jin_SciRep_24}. We have not experimentally demonstrated this phenomenon, as we find kicks any larger than those we have applied herein, exceed the pull-in voltage of our device, leading to an irreversible collapse of the membrane. We anticipate that the same experimental protocol could be applied to devices with larger membrane-substrate separations to demonstrate full dynamic decoupling.

\begin{figure}[!h]
    \begin{center}
    \includegraphics[width= 1 \linewidth]{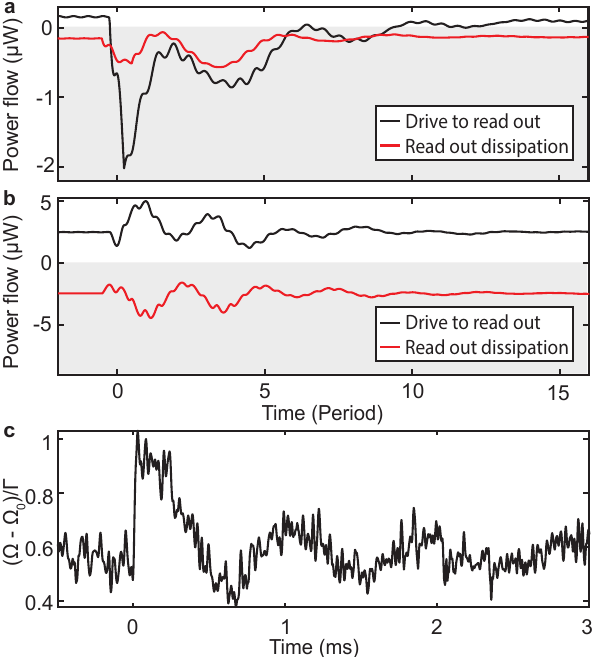}
    \end{center}
    \caption{\textbf{Error correction mechanism.} \textbf{a/b} Power flow from the external drive to the read out resonator (black) and the intrinsic dissipation of the read out resonator (red), with the system initialised in the `0'/`1' state. The direction of the power flow is defined such that positive values correspond to the read out resonator gaining energy, and negative values (gray background) indicate a loss of energy. The x-axis is normalised to dimensionless units, representing the number of oscillation periods, with $t=0$ being the time of the kick. \textbf{c} The frequency shift of the read out resonator as a kick is applied at $t=0$ ms. The y-axis represents the shift away from the resonant frequency of the resonator, in terms of linewidths.}
    \label{error correction mechanism}
\end{figure}

\subsection*{Error Correction Statistics}
To quantify the effectiveness of the coupled system at performing error correction, we carry out repeated measurements to determine the probability of `0' to `1' and `1' to `0' errors. These results are benchmarked against analogous experiments on a single resonator. We determine the `0' to `1' ($P_{0 \to 1}$) and `1' to `0' ($P_{1 \to 0}$) failure probabilities separately as they are not necessarily equal. Indeed, the probabilities of each type of error are predicted to vary depending on the frequency and strength of the AC drive \cite{Jin_SciRep_24}. When the drive is strong or is closer to the resonant frequency, the system can more easily transition from low to high amplitude. Under these conditions, the system will be more prone to `0' to `1' errors. In contrast, when the drive is weak or far away from resonance, the system will be more prone to `1' to `0' errors. 

A total of 1295 kicks were performed, with the resulting failure probabilities and 95$\%$ confidence intervals shown in Fig. \ref{Error correction map}a. (see Supplementary Information for data). For a single resonator, we observe frequent bit flip errors and record $P_{1 \to 0}$ and $P_{0 \to 1}$ values of 0.35 and 0.15, respectively. The difference between these two failure probabilities are due to the single resonator being driven far from resonance. In comparison, when operating at the same condition, the coupled resonators consistently suppress both types of errors, with $P_{1 \to 0}$ and $P_{0 \to 1}$ values of 0.0065 and 0.0060, respectively.

In total, we calculate the overall failure probability ($P_\text{fail}$) to be 0.22 for the single resonator with a 95$\%$ confidence interval of 0.19 - 0.25, and 0.0063 for the coupled system with a 95$\%$ confidence interval of 0.0008 - 0.022. This corresponds to a 35-fold error rate reduction.

Previous theoretical work suggested that the coupled system should fully suppress all single event upsets \cite{Jin_SciRep_24}. To explore the source of errors in our experiments, we extend the simulations in our theoretical work \cite{Jin_SciRep_24} to include thermal noise as well as parameter mismatch and cross-talk between the resonators (see Supplementary Information). We find that the system is robust to both thermal noise and parameter mismatch over the range that is possible in our experiments. On the other hand, our simulations show that the system is sensitive to cross-talk at the levels that are experimentally present. This cross-talk has the effect of simultaneously altering the amplitude, frequency and phase of the ancilla resonators when the read out resonator is kicked. At specific kick phases, this cumulative effect across all resonators causes the observed errors that would otherwise be corrected. Other environmental factors such as temperature change can cause gradual shifts in resonant frequency and hence result in transitions between steady states. However, we rule this out on the basis that our devices are temperature controlled and we do not observe anomalous windows of bit flips.

\begin{figure}[!b]
\begin{center}
\includegraphics[width= 1 \linewidth]{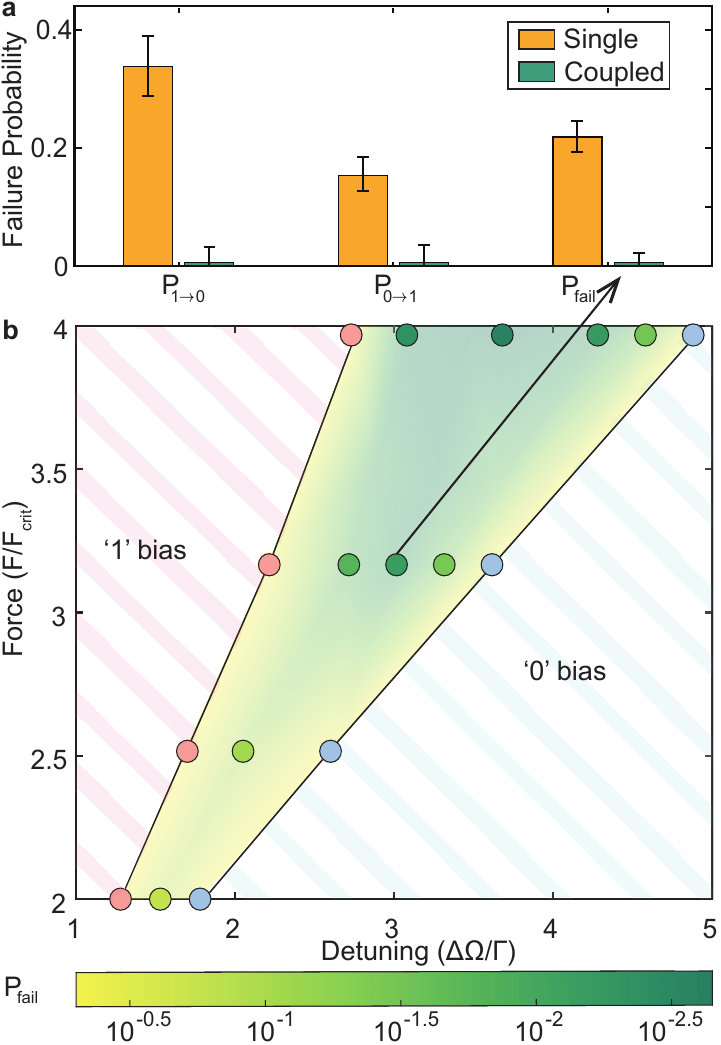}
\end{center}
\caption{\textbf{Error correction phase map.} \textbf{a} Failure probability statistics for a single resonator and three coupled resonators, measured at $(\Delta \Omega /\Gamma, F/F_{\text{crit}})  = (3.0,3.2)$. Here, $\Delta \Omega$ is the difference between the drive and the resonant frequency, $ F$ is the drive strength and $F_\text{crit}$ is the critical drive. $P_{0 \to 1}$ and $P_{1 \to 0}$ respectively denote the `0' to `1' failure probability and the `1' to `0' failure probability. $P_\text{fail}$ denotes the overall failure probability. The $95 \%$ confidence intervals on the failure probabilities, as determined via the Clopper-Pearson method \cite{Clopper_34_Biometrika}, are displayed by the error bars. \textbf{b} Experimental map showing the mean probability of failure at various drive strengths and detunings. The interpolated background is fainted so as to not obscure the data points. The boundary of the Duffing bistability regions are plotted as red and blue dots. Within the `1'/`0' bias region, the resonators can only evolve into their collective `1' state or their collective `0' state.} 
\label{Error correction map}
\end{figure}

\subsection*{Error Correction Map}

To validate the prediction of an error correcting phase, we test the performance of our device at different points in the parameter space. Given that the error correcting phase can only exist within the bistable regime, we first performed forward and backward frequency sweeps at various drive strengths to determine the extent of this region. The forward sweep allows us to determine the higher frequency boundary of the Duffing hysteresis, while the backward sweep reveals the lower frequency boundary \cite{Schmid, Romero_PRA_24}. Our measurements reveal a wedge-shaped bistability region that shifts towards higher frequencies and covers a broader range of detunings with increasing drive strength, as shown in Fig. \ref{Error correction map}b. This is in good qualitative agreement with our theoretical prediction (see Supplementary Information).

Within this region, we then measured the response of the read out resonator to hundreds of randomly phased kicks at each of nine different detuning and drive strength operating points (2988 kicks in total, see Supplementary Information for detailed statistics). We calculate the overall probability of failure of the device at each operating point, shown as color coded data points in Fig. \ref{Error correction map}b. We find that the error correcting performance of the coupled device improves with increasing drive force. Notably, at the center of the bistable region, $(\Delta \Omega /\Gamma, F/F_{\text{crit}})  = (3.7,3.9)$, we record the failure probability to be as low as $0.0020$, corresponding to just one bit flip error observed from a total of 495 applied kicks.

As predicted in theoretical work \cite{Jin_SciRep_24}, the error correction functions over a wide region of drive detunings and forces within the bistable region, evidencing the robustness of the scheme to parameter drifts. As shown in the background of Fig. \ref{Error correction map}b, a nearest-neighbour interpolation of our data reveals a wide area of low failure probabilities. For instance, at $F/F_{\text{crit}} = 3.9$, the device supports error correction over a detuning range of 1.2 linewidths. Similarly, at $\Delta \Omega/\Gamma = 3.0$, the system can perform error correction over a force range as wide as 0.7 $F_{\text{crit}}$.

\section*{Discussion}

This work experimentally demonstrates a new error correction scheme, reliant on the emergence of an \textit{error correcting phase} wherein corrupted memory states are autonomously restored via the collective dynamics of three coupled nonlinear resonators. The approach is versatile and can be implemented in a broad range of architectures that host nonlinear resonators, such as electrical, optical \cite{Moss}, superconducting \cite{KounalakisM_18_nature_comm} and nanomechanical circuits \cite{Wenzler_Nano_Lett_14}. 

Our work focuses on nanomechanical computing since it is an approach that promises radiation-robust and ultra low energy consumption, potentially approaching the Landauer limit \cite{Lee_NatComm_23, Romero_PRA_24}, and because purely nanomechanical error correction has not previously been realised. A significant advantage of our approach is that the emergent collective dynamics of the nonlinear system enable passive, autonomous error correction. This is in contrast to conventional majority-voting algorithms, which typically require continuous reading and rewriting of data bits \cite{Shimeall_14_book}. By showing that nanomechanical error correction is possible, our work contributes to the realisation of a scalable nanomechanical processor.

It is interesting to consider the potential causes of errors within a nanomechanical computer. Single-event upsets can be induced by electromagnetic pulses, ionising radiation, or mechanical shock \cite{Lee_NatComm_23}. High energy ionising particles are of particular concern due to their ability to penetrate materials, with studies estimating that radiation causes an average of twelve bit flip errors per day in low-Earth orbit electronics \cite{Alexander_18_IEEE}. These errors are typically triggered by energy depositions in the 10 – 100 keV range \cite{Wilcox_21_IEEE}, directly comparable to our experiments where each kick delivers approximately 70 keV of energy to the resonator (see Supplementary Information). Moreover, it has been predicted that the dynamic decoupling which we demonstrate, can enable the correction of even higher energy impulses \cite{Jin_SciRep_24}. This suggests that the scheme is capable of correcting impulses typical of a realistic operating environment.

Beyond error correction, our work highlights that even a minimal network of three coupled nonlinear resonators can exhibit complex dynamics. Our platform offers high quality factors, strong nonlinearity and tunable mechanical coupling, and can be readily scaled to larger arrays to enable systematic studies of driven, dissipative nonlinear systems \cite{Romero_PRA_24}. Such systems are known to host rich collective behaviour, with up to $3^N$ stable states for a system of N elements \cite{MargianiGabriel2025TscK}. For example, it has been demonstrated that eight coupled resonators exhibit collective phenomena including weak chimeras, decoupled states, travelling waves, inhomogeneous synchronised states and symmetry breaking \cite{Roukes_science_2019}. They are also important for a range of emerging technologies such as novel frequency comb states \cite{Ochs_22_PRX, Ganesan_17_PRL}, Ising machines for ultra-fast optimisation \cite{McMahon_16_Science, Heugel_22_PRR, Lee_25_Nature_Comm}, and energy-efficient neuromorphic computers \cite{Chen_25_science}. The scalability and high nonlinearity of our system make it well-suited for exploring these sorts of collective nonlinear dynamics, both for their rich physics and in pursuit of technological advances.

\section*{Methods}

\subsection*{Fabrication}
The devices are fabricated from chips diced from a commercially available wafer (Microchemicals GmbH). The wafer consists of three layers - silicon nitride ($50 $ nm) deposited on top of a sacrificial silicon oxide layer ($500 $ nm), on top of a silicon substrate (500 $\mu$m). An additional layer of gold electrode (45 nm) with a chromium adhesion layer (5 nm) is patterned by electron-beam lithography, and then deposited using an electron-beam evaporator. The silicon nitride membranes are patterned with 1 $\mu $m by 1 $\mu$m holes using electron-beam lithography and reactive ion etching using CHF$_\text{3}$, CF$_\text{4}$, and O$_\text{2}$ in the ratio 25:40:4 standard cubic centimetres per minute. These holes allow for a top down release process, wherein the silicon oxide is etched away by immersing the chip in buffered hydrofluoric acid \cite{Hirsch_24_APL}. Finally, the chips are dried using a critical point dryer and wire bonded to printed circuit boards.

\subsection*{Optical read out}

We optically measure the response of the read out resonator, as shown in Fig. 1b. Heterodyne detection relies on the interference of a weak probe signal and a strong frequency shifted field, typically referred to as the local oscillator (LO). The probe and the LO electric fields, $E_p$ and $E_{LO}$, respectively be described as: \begin{align*}
    E_p &= A_p \exp{(i (\omega + \omega_\text{LO}) t + \phi_L + \phi_s)}\\
    E_{LO} &= A_{LO} \exp{(i \omega t)}
\end{align*}
Here, $A_p$/$A_{LO}$ is the amplitude of the probe/LO field, $\omega$ is the laser frequency, and $\omega_\text{LO}$ is the frequency shift created by the AOM (77 MHz). $\phi_s$ represents the phase change introduced by the motion of the membrane, and $\phi_L$ represents the relative phase change between the two interferometry arms. Since the resonator is driven with a sinusoidal signal at the resonant frequency, $\phi_s$ can be expressed as $\phi_s = A_m \cos (\Omega_D t)$, where $A_m$ is the amplitude of the phase modulation. If $A_m = 0$, the membrane has zero motion. If $A_m = 2 \pi$, the membrane's vibrational amplitude corresponds to the laser wavelength (780 nm in this case).

As the probe and the LO fields interfere, the output of the photocurrent detector produces a signal at the AOM frequency ($i_\text{AOM}$), and two sidebands signals ($i_\text{mech}$) at $ \omega_\text{LO} + \Omega_0$ and $ \omega_\text{LO} - \Omega_0$. To a good approximation, these signals can be described by:
\begin{align*}
i_\text{AOM} & \approx A_{LO} A_p \cos ( \omega_\text{LO} t + \phi_L) \\
i_\text{mech} & \approx -2 A_{LO} A_p A_m \sin ( \omega_\text{LO} t + \phi_L) \cos (\Omega t)\\
  &= - A_{LO} A_p A_m ( \sin(( \omega_\text{LO} + \Omega_0)t + \phi_L) \\
  & +  \sin(( \omega_\text{LO} - \Omega_0)t + \phi_L) 
\end{align*}

Using the above expressions, one can extract $A_m$ by taking the ratio of the power at the sideband frequency and the AOM frequency. $A_m$ can then be converted from radians to meters using the expression $x = A_m \cdot \frac{780 }{2 \pi}$ nm. 

\subsection*{Electrostatic actuation}

The gold electrodes deposited on top of the resonators allow the devices to be actuated capacitively. To a good approximation, the gold electrode and the substrate forms a parallel-plate capacitor, with the capacitance ($C$) given by $C = \varepsilon_0 A/(d_0 + u_0)$ \cite{Romero_PRA_24}. Here, $\varepsilon_0$ is the vaccum permittivity constant, $A$ is the area of the electrode, $d_0$ is the plate separation when the membrane is not deformed, and $u_0$ is the displacement at the center of the resonator. The capacitive force applied to the membrane, is then given by:\begin{align*}
    F_C &= \frac{ \varepsilon_0 A}{2(d_0 + u_0)^2}V^2
\end{align*}

Since the force scales with the voltage applied squared, only applying an AC signal near mechanical resonance ($V_{AC} \propto \sin (\Omega_0 t)$) to the electrode would lead to a response at $2 \Omega_0$. In comparison, when simultaneously applying an AC signal and a DC signal ($V_{DC}$) to the electrode, the above expression can be expanded into: \begin{align*}
    F_C &= \frac{ \varepsilon_0 A (V_{DC}^2 + 2 V_{AC} V_{DC} + V_{AC}^2) }{2(d_0 + u_0)^2}
\end{align*}

This allows us to actuate motion at $\Omega_0$ with the response boosted by a factor of $2 V_{DC}/V_{AC}$.

\subsection*{Phase read out}

The phase of the mechanical oscillation can be measured via postprocessing the recorded photocurrent (optical path shown in Fig. 2b). Post processing is required because the relative phase between the two interferometry arms is not stabilised, meaning that the phase of the photocurrent signal randomly drifts over time \cite{Ralph_quantum_optics_textbook}, interfering with the measurement.


Post processing is performed by passing the photocurrent through two RF mixers. The LO ports of the two mixers were provided with the AOM signal, and a $\pi/2$ phase shifted AOM signal. We refer to the output of the mixers respectively as $i_X$ and $i_Y$. Finally, $i_X$ and $i_Y$ pass through low pass filters so that only the down-converted signals are measured. The phase drift between the interferometry arms can then be calculated using the formula:
\begin{align*}
     \phi_L(t) = \displaystyle \arctan\left(\frac{i_Y}{i_X}\right)
\end{align*}
Knowing $\phi_L(t)$, we can then correct the photocurrent to remove any phase drift:
\begin{align*}
    i_{X'} = \cos (\phi_L)  i_X + \sin (\phi_L)  i_Y\\
    i_{Y'} = \sin (\phi_L)  i_X - \cos (\phi_L)  i_Y
\end{align*}

To track the evolution of the resonator in phase space in a frame rotating with the input sinusoidal drive, we plot $i_{X'} \times \sin (\Omega_D t)$ against $i_{Y'} \times \sin (\Omega_D t)$, where $\Omega_D$ is the drive frequency. By normalising both axes with $x_\text{crit}$, we obtain the experimental results shown in Fig. 2 and Fig. 4. 

\subsection*{Simulating resonator dynamics}
The single resonator case is described by the ODE for a driven, damped Duffing oscillator \cite{Schmid, Romero_PRA_24}:
\begin{align*} 
    \ddot{x} +  \Gamma \dot{x} +  \Omega_0^{2} x +  \frac{\alpha}{m} x^3 &= \frac{F}{m} \cos(\Omega_D t)
\end{align*}
where  $x$ is the displacement, $m$ is the mass, $\Gamma$ is the dissipation, $\Omega_0$ is the resonant frequency, $\alpha$ is the nonlinear Duffing coefficient, and $F$ is the amplitude of the sinusoidal drive provided to the resonator at frequency $\Omega_D$. Driven, damped Duffing resonators are known to exhibit complex dynamics, and there exists no closed form analytical solution. In this work, to simulate a single mechanical memory unit and its response to kicks,  we numerically solve the above Duffing equation using an ODE solver. 

In Fig. 2b, a single resonator is initialised with `0' and `1' state by providing the ODE solver with a low or high initial displacement. Kicks are introduced by changing the resonator's velocity between simulation timesteps, which physically corresponds to an instantaneous change in momentum, denoted as $\Delta p$. The parameters used to generate Fig. 2b are: $m = 10^{-12}$ kg, $\Gamma = 10^{5}$ s$^{-1}$, $\omega_0 = 10^{6}$ s$^{-1}$, $\alpha/m = 3 \times 10^{22}$ m$^{-2} s^{-2}$, $\omega = 1.152 \times 10^{6}$ s$^{-1}$, $F  = 5 \times 10^{-7}$ N, $\Delta p = 5.5 \times 10^{-12}$ kg.m.s$^{-1}$.

Similarly, three all-to-all coupled resonators (equivalent to the physical system shown in Fig. 3b) can be described by the below equations of motion:
\begin{align*}
 \ddot{x}_1 + \Gamma \dot{x}_1 + \Omega_{0}^2 x_1 + \frac{\alpha}{m} x_1^{3}& = \frac{F}{m} \cos(\Omega_D t) + \beta x_2 + \beta x_3  \\
\ddot{x}_2 + \Gamma \dot{x}_2 +  \Omega_0^2 x_2 + \frac{\alpha}{m} x_2^{3}  & =\frac{F}{m} \cos(\Omega_D t)  + \beta x_1 + \beta x_3  \\
 \ddot{x}_3 +  \Gamma \dot{x}_3 +  \Omega_0^2 x_3+ \frac{\alpha}{m} x_3^{3}  &= \frac{F}{m}   \cos(\Omega_D t)  + \beta x_1 + \beta x_2  
\end{align*} 
where $\beta$ is the coupling constant between resonators, and $x_1$, $x_2$, and $x_3$ represent the displacement of resonators $1$, $2$ and $3$, respectively. The phase trajectories of the coupled system shown in Fig. 4a and 4b were generated by solving the above equations, with a kick introduced on one of the resonators. The parameters used in Fig. 4 are given by:  $m = 10^{-12}$ kg, $\Gamma = 10^{5}$ s$^{-1}$, $\omega_0 = 10^{6}$ s$^{-1}$ , $\alpha/m = 3 \times 10^{22}$ m$^{-2}.s^{-2}$, $\omega = 1.152 \times 10^{6}$ s$^{-1}$, $F=1.048 \times 10^{-6}$ N, $\Delta p = 4 \times 10^{-12}$ kg.m.s$^{-1}$, $\beta = 2 \times 10^{11}$ s$^{-2}$.

\subsection*{Simulating power spectrum of coupled resonators}

When our fabricated resonators are weakly driven ($F < F_\text{crit}$), they behave as linear resonators. The mechanical power spectrum of coupled, linear resonators can be analytically solved as follows. 

The behaviour of our resonators under these conditions is governed by the below equations:
\begin{align*}
 m \ddot{x}_1 + m \Gamma \dot{x}_1 + m \Omega_{1}^2 x_1   - m \beta x_2 - m \beta x_3 &= F_1    \\
m \ddot{x}_2 + m \Gamma \dot{x}_2 +  m \Omega_2^2 x_2  - m \beta x_1 - m \beta x_3  & =F_2    \\
 m \ddot{x}_3 + m \Gamma \dot{x}_3 +  m \Omega_3^2 x_3 - m \beta x_1 - m \beta x_2  &= F_3     
\end{align*} 
where $\Omega_{1,2,3}$ are the resonant frequencies of the three resonators. We assume that the above equations have a solution in the form of $\mathbf{x}(t) = \mathbf{A} \exp{i \nu t} $, where $\mathbf{x} = [x_1, x_2, x_3]^T$ are the displacements, $\mathbf{A} = [A_1, A_2, A_3]^T$ are the peak to peak oscillation amplitudes, and $\nu$ is the oscillation frequency. We also assume that the force is of the form $\mathbf{F}(t) = \mathbf{f} \exp(i \nu t)$, with $\mathbf{x} = [f_1, f_2, f_3]^T$.

Then, the above three equations can be represented as:\begin{align*}
- \nu^2 \mathbf{M} \mathbf{A} + i \Gamma \nu \mathbf{M} \mathbf{A} + \mathbf{K} \mathbf{A} & = \mathbf{f}
\end{align*} 
where,
\begin{equation*}
\mathbf{M} = \begin{pmatrix} m & 0 & 0 \\ 
0 &m & 0 \\
0 &0 & m  \end{pmatrix} 
\end{equation*}

\begin{equation*}
 \mathbf{K}=\begin{pmatrix} m \Omega_1^2 & - m \beta & - m \beta \\ 
  - m \beta & m \Omega_2^2 & - m \beta \\ 
- m \beta & - m \beta & m \Omega_3^2  \\ 
\end{pmatrix} 
\end{equation*}

The solution to the above equation is given by:
\begin{align*}
    \mathbf{A}(\nu) = [- \nu^2 \mathbf{M} + i \Gamma \nu \mathbf{M} + \mathbf{K} ]^{-1} \mathbf{f}
\end{align*}

The background mechanical power spectrum in Fig. \ref{fig:coupled_schematic}a is generated by solving $\mathbf{A}(\nu)$ while varying $\Omega_3$, which corresponds to experimentally applying DC voltages to resonator 3. 

The following values were used: $m = 7.4728 \times 10^{-12}$ kg, $\Gamma/2 \pi = 175 $ s$^{-1}$, $\Omega_1/2 \pi = 4.131650 \times 10^6$ Hz, $\Omega_2/2 \pi = 4.141050 \times 10^6$ Hz, $\Omega_3/2 \pi = 4.142250 \times 10^6$ Hz, $f_1 = 0$ N, $f_2 = 0$ N, $f_3 = 1.1121 \times 10^{-11}$ N, $\beta =3459 s^{-1}$. The critical force of the above system is given by $F_\text{crit} = 2.172 \times 10^{-10}$ N, which is much larger than the value of $f_3$.

\section*{Data Availability}
All data needed to evaluate the conclusions in the paper are present in the paper and/or the Supplementary Information. Additional simulation files and scripts are accessible upon request from W.P.B.

\providecommand{\noopsort}[1]{}\providecommand{\singleletter}[1]{#1}%

\section*{Acknowledgements}

This research was primarily funded by the Australian Research Council and Lockheed Martin Corporation through the Australian Research Council Linkage Grant No. LP190101159. Support was also provided by the Australian Research Council Centre of Excellence for Engineered Quantum Systems (No. CE170100009). G.I.H. (No. DE210100848) and C.G.B. (No. FT240100405) acknowledge their Australian Research Council grants.

\section*{Author contributions}
X.J: Conceptualisation, numerical simulation, finite element simulation, methodology, fabrication, data collection, data analysis, writing.\\
C.G.B: Conceptualisation, methodology, data analysis, editing, supervision, funding acquisition.\\
N.A: Finite element simulation, fabrication, editing.\\
N.P.M: Conceptualisation, finite element simulation, methodology, data analysis, editing.\\
E.R: Methodology, fabrication.\\
T.M.F.H: Methodology, fabrication.\\
G.I.H:  Conceptualisation, methodology, data analysis, editing, supervision, funding acquisition.\\
W.P.B: Conceptualisation, methodology, data analysis, editing, supervision, funding acquisition.\\
All authors reviewed the manuscript. 

\section*{Competing interests}
The authors declare no competing interests.

\end{document}